\documentstyle[12pt]{article}
\title{A linear realization for the new space-time superalgebras
in ten and eleven dimensions}
\author{Alexei Deriglazov and Anton Galajinsky}
\date{Department of Mathematics and Mathematical Physics,
\thanks{deriglaz@phys.tsu.tomsk.su, galajin@phys.tsu.tomsk.su\/}\\
Tomsk Polytechnical University, 634004 Tomsk, Russia}
\begin{document}
\maketitle
\large
\begin{abstract}
The new extensions of the Poincar\'e superalgebra recently found in
ten and eleven dimensions are shown to admit a linear realization.
The generators of the nonlinear and linear group transformations are
shown to fall into equivalent representations of the superalgebra.
The parametrization of the coset space $G/H$, with $G$ a given extended
supergroup and $H$ the Lorentz subgroup, that corresponds to the linear
transformations is presented.
\end{abstract}

\section{Introduction}
  Recently, new extensions of the  Poincar\'e superalgebra were found in
ten and eleven dimensions [1--5]. To a large extent they can be thought
of as suggesting a natural geometric framework for extended
objects [6--8].
A covariant formulation of the super $p$-brane theory essentially
involves a Wess--Zumino-type term. Being crucial in providing local
$k$-symmetry, this term is invariant under global supersymmetry
transformations up to total derivative only. As was shown in Ref. 9,
this causes the effect of introducing a topological central charge
into the Poincar\'e superalgebra (the effect also known for the massive
superparticle in $d=9$ [10]). The total derivative terms in a
variation of the action can be suppressed by redefining the action
functional itself (i.e., by enlarging the original configuration space
and adding appropriate total derivative terms to the action). It turns
out that following this course one arrives at an extension of the
Poincar\'e superalgebra again [11, 3].

A conventional way to build a theory associated with a given Lie
(super) algebra is to apply the standard group-theoretic construction
[12]. For the case concerned, it is suffice to consider the coset
space $G/H$, where $G$ is a given extended supergroup , with an
element
\begin{equation}
\tilde g = e^{-ia^nP_n+i\epsilon^\alpha Q_\alpha+i\sigma_{\cdots}
\Sigma^{\cdots}+\frac i2\omega^{mn}M_{mn}},
\end{equation}
and $H$ is the Lorentz subgroup. The symbol
$i\sigma_{\cdots}\Sigma^{\cdots}$ denotes the sum over all generators
$\Sigma^{\cdots}$ that extend the Poincar\'e superalgebra $(M,P,Q)$.
A point in the space
\begin{equation}
g(x,\theta,\psi)=e^{-ix^nP_n+i\theta^\alpha Q_\alpha+i\psi_{\cdots}
\Sigma^{\cdots}}\times SO(1,d-1)
\end{equation}
is parametrized by the set of coordinates $(x^m,\theta^\alpha,\psi_{\dots})$
with the statistics being analogous to that of the generators.

Left multiplication with a group element
\begin{equation}
\tilde g:\quad g(x,\theta,\psi)\to g(x',\theta',\psi')=\tilde g
g(x,\theta,\psi)
\end{equation}
defines, via the Baker--Campbell--Haussdorff formula, an action of the
group on the coset. Invariants of the group can be used to construct a
theory.

A salient feature of the new superalgebras is that the translation
generators do not commute with the supertranslations [1--4]
\begin{equation}
[P,Q]\sim\Sigma.
\end{equation}
In view of the construction just outlined it means highly nonlinear
transformation laws for the coordinates parametrizing the coset.

The purpose of this letter is to show that the extended superalgebras
recently proposed in [1--3] admit a linear realization. The
$\Gamma$-matrix identities that underlie the superstring theory in
$d=10$ and the supermembrane theory in $d=11$ turn out to be important
for the linearization.

Apart from the obvious technical benefits, there is an additional
motivation to study the extended superalgebras in linear realization.
The important observation is that, due to Eq. (4), the first Casimir
operator in the algebra (see Eqs. (17) and (12) below) includes the
odd generators contribution. This allows one to expect that equations
extracting on-shell field irreps of the group will involve interaction.
Construction of the irreps may suggest an interesting way to attack
the higher spins interaction problem [13]. The results on this subject
will be present elsewhere.

In this work we use (anti)symmetrization ``without strength'', i.e.,
$A_{[m}B_{n]}\equiv A_mB_n-A_nB_m$, $A_{(m}B_{n)}\equiv A_mB_n+A_nB_m$.
The conventions adopted for $d=11$ are presented in the Appendix.

\section{Green superalgebra}
  Green superalgebra [1] can be
understood as a global limit of Ka\v{c}--Moody one [14, 2] that
arises in the context of superstring theory. The commutation relations
read (the usual Poincar\'e subalgebra is omitted)
\begin{equation}
\begin{array}{ll}
\{Q_\alpha,Q_\beta\}=-2{\Gamma^m}_{\alpha\beta}P_m, &
[P_m,Q_\alpha]=-i\Gamma_{m\,\alpha\beta}\Sigma^\beta,\\
{}[M_{mn},Q_\alpha]=-\displaystyle\frac i4(\Gamma_{mn})_\alpha{}^\beta
Q_\beta, & [M_{mn},\Sigma^\alpha]=\frac i4 \Sigma^\beta
(\Gamma_{mn})_\beta{}^\alpha, \end{array}
\end{equation}
where $\Gamma_{mn}\equiv \Gamma_m\tilde\Gamma_n-\Gamma_n\tilde\Gamma_m$
and $\Gamma_m\tilde\Gamma_n+\Gamma_n\tilde\Gamma_m=-2\eta_{mn}$. The
Jacobi identities for Eq. (5) restrict the $\Gamma$-matrices to satisfy
the relation
\begin{equation}
{\Gamma^m}_{\alpha(\beta}\Gamma_{m\,\gamma\delta)}=0
\end{equation}
which holds only in $d=3,4,6$ and 10. For definiteness we examine
ten-dimensional case here\footnote{The conventions adopted in $d=10$
are those of Ref. 15.}. The generalization to other dimensions is
straightforward.

An application of the group-theoretic construction to the case at
hand results in the transformation laws [1, 3]
$$\begin{array}{l} \delta_\epsilon\theta^\alpha=\epsilon^\alpha,\qquad
\delta_\epsilon x^m=i\theta\Gamma^m\epsilon,\\
\delta_\epsilon\psi_\alpha=\displaystyle\frac 12 x^m
(\epsilon\Gamma_m)_\alpha-\frac 16 i(\theta\Gamma^m\epsilon)
(\theta\Gamma_m)_\alpha;\end{array}
\eqno{(7a)}$$
$$
\delta_ax^m=a^m,\qquad \delta_a\psi_\alpha=-\frac 12 a^m
(\theta\Gamma_m)_\alpha;
\eqno{(7b)}$$
$$\begin{array}{l} \delta_\omega x^m={\omega^m}_nx^n,\qquad
\delta_\omega\theta^\alpha=\displaystyle\frac 18 \omega^{mn}
(\theta\Gamma_{mn})^\alpha,\\
\delta_\omega\psi_\alpha=-\displaystyle\frac 18\omega^{mn}
(\Gamma_{mn}\psi)_\alpha;\end{array}
\eqno{(7c)}$$
$$
\delta_\sigma\psi_\alpha=\sigma_\alpha,
\eqno{(7d)}$$
where $(x^m,\theta^\alpha,\psi_\beta)$ are the coordinates
parametrizing the coset $G/H\sim {R}^{10/32}$ and $(\epsilon, a,
\omega,\sigma)$ are the parameters associated to the generators
\addtocounter{equation}{1}
\begin{equation}
\begin{array}{l}
Q_\alpha=i\partial_\alpha+(\theta\Gamma^n)_\alpha\partial_n+
\displaystyle\frac i2x^n\Gamma_{n\,\alpha\beta}\partial^\beta-\frac 16
(\theta\Gamma^n)_\alpha(\theta\Gamma_n)_\beta\partial^\beta,\\
P_n=-i\partial_n+\displaystyle\frac i2(\theta\Gamma_n)_\alpha
\partial^\alpha,\\
M_{mn}=-i(x_m\partial_n-x_n\partial_m)+\displaystyle\frac i4
(\theta\Gamma_{mn})^\alpha\partial_\alpha-\frac i4
(\Gamma_{mn}\psi)_\alpha\partial^\alpha,\\
\Sigma^\alpha=i\partial^\alpha.\end{array}
\end{equation}
In Eq. (8) we have set $\partial_\alpha\equiv\partial/\partial
\theta^\alpha$, $\partial^\alpha\equiv\partial/\partial\psi_\alpha$,
$\partial_n\equiv\partial/\partial x^n$. Note that Eq. (7a) essentially
involves a nonlinear contribution.

In terms of the transformations (7) the algebra (5) acquires the form
\begin{equation}
\begin{array}{ll}
[\delta_{\epsilon_1},\delta_{\epsilon_2}]=\delta_a, & a^m=2i\epsilon_1
\Gamma^m\epsilon_2,\\
{}[\delta_a,\delta_\epsilon]=\delta_\sigma, & \sigma_\alpha =a^m
(\epsilon\Gamma_m)_\alpha,\\
{}[\delta_\omega,\delta_\epsilon]=\delta_{\epsilon_1}, &
\epsilon_1^\alpha=-\displaystyle\frac 18 \omega^{ab}
(\epsilon\Gamma_{ab})^\alpha,\\
{}[\delta_\omega,\delta_\sigma]=\delta_{\sigma_1}, &
\sigma_{1\,\alpha}=\displaystyle\frac 18 \omega^{ab}
(\Gamma_{ab}\sigma)_\alpha,\\
{}[\delta_\omega,\delta_a]=\delta_{a_1}, & {a_1}^m=-{\omega^m}_na^n,\\
{}[\delta_{\omega_1},\delta_{\omega_2}]=\delta_{\omega_3}, &
{\omega_3}^m{}_n=(\omega_2\omega_1)^m{}_n-(\omega_1\omega_2)^m{}_n.
\end{array}
\end{equation}
Evaluation of this algebra turns out to be more instructive than it
may seen at a glance.. Actually, direct calculation with the use of
Eq. (6) yields
\begin{equation}
\begin{array}{l} \delta_{\epsilon_1}\Big(\displaystyle\frac 12 x^n
(\epsilon_2\Gamma_n)_\alpha\Big)-\delta_{\epsilon_2}\Big(\frac 12 x^n
(\epsilon_1\Gamma_n)_\alpha\Big)=-\frac 12i(\epsilon_1\Gamma^n
\epsilon_2)(\theta\Gamma_n)_\alpha,\\
\delta_{\epsilon_1}\Big(-\displaystyle\frac 16 i(\theta\Gamma^n
\epsilon_2)(\theta\Gamma_n)_\alpha\Big)-\delta_{\epsilon_2}\Big(
-\frac 16 i(\theta\Gamma^n\epsilon_1)(\theta\Gamma_n)_\alpha\Big)=-\frac
12 i(\epsilon_1\Gamma^n\epsilon_2)(\theta\Gamma_n)_\alpha,\end{array}
\end{equation}
which means that the linear and nonlinear terms in the variation
$\delta_\epsilon\psi$ make the same contribution into the first line
of Eq. (9). The latter fact implies that Eq.(7) can be rewritten in
the linear form without spoiling the algebra (9). The linear version for
the Green transformations looks like
\begin{eqnarray}
&& \delta_\epsilon\theta^\alpha=\epsilon^\alpha,\qquad
\delta_\epsilon x^m=i\theta\Gamma^m\epsilon,\qquad
\delta_\epsilon\psi_\alpha=\displaystyle\frac 23 x^m
(\epsilon\Gamma_m)_\alpha;\cr
&& \delta_ax^m=a^m, \qquad \delta_a\psi_\alpha=-\displaystyle\frac 13
a^m(\theta\Gamma_m)_\alpha;\\
&& \delta_\sigma\psi_\alpha=\sigma_\alpha,\nonumber
\end{eqnarray}
where we omitted the usual Lorentz transformations. The generators
associated to Eq. (11) read
\begin{eqnarray}
&& Q_\alpha=i\partial_\alpha+(\theta\Gamma^n)_\alpha\partial_n+
\displaystyle\frac 23 ix^n\Gamma_{n\,\alpha\beta}\partial^\beta,\cr
&& P_n=-i\partial_n+\displaystyle\frac i3(\theta\Gamma_n)_\alpha
\partial^\alpha,\\
&& \Sigma^\alpha=i\partial^\alpha.\nonumber
\end{eqnarray}

In geometric terms, a realization of the algebra (5) by the linear
transformations corresponds to the possibility to choose another
parametrization of the coset (in this context see also Ref. 16). It is
straightforward to check that adopting the following parametrization:
\begin{equation}
g=e^{-ix^nP_n+i\psi_\alpha\Sigma^\alpha+\frac 23 i\theta^\alpha
Q_\alpha} e^{\frac 13i\theta^\alpha Q_\alpha}\times SO(1,d-1)
\end{equation}
and exploiting the group-theoretic machinery one arrives just at Eq.
(11). It is interesting to note that the generators (8) and (12) fall
into two equivalent representations of Green superalgebra\footnote{The
explicit form for the operator $S$ immediately follows from a comparison
of Eqs. (2) and (13) with the latter being rewritten in the equivalent form
$g=e^{-ix^nP_n+i\psi_\alpha\Sigma^\alpha+\frac 23 i\theta^\alpha Q_\alpha}
e^{\frac 13 i\theta^\alpha Q_\alpha}\times SO(1,d-1)=e^{-ix^nP_n+
i\theta^\alpha Q_\alpha+i\left(\psi_\alpha -\frac 16 x^n
(\theta\Gamma_n)_\alpha \right)\Sigma^\alpha}\times SO(1,d-1)$.}
\begin{equation}
T_{i\rm(nonlinear)}=ST_{i\rm(linear)}S^{-1},
\end{equation}
where
\begin{equation}
S=e^{\frac 16 x^m(\theta\Gamma_m)_\alpha\partial^\alpha},
\end{equation}
and we denoted $T_i\equiv (P,Q,M,\Sigma)$. This relation can easily
be checked by making use of the formula
\begin{equation}
\begin{array}{c} e^{-B}Ae^B =\sum\limits_{n=0}^\infty\displaystyle\frac 1{n!}
[A,B]_{(n)},\\[2ex]
{}[A,B]_{(0)}\equiv A, \qquad [A,B]_{(n+1)}=[[A,B]_{(n)},B].\end{array}
\end{equation}
Thus, Green superalgebra was shown to admit a linear realization.

Some remarks seem to be relevant here. First, since momenta do not
commute with supertranslations the first Casimir operator in the
algebra includes the odd generators contribution
\begin{equation}
P^2-i\Sigma^\alpha Q_\alpha.
\end{equation}
This allows one to expect that equations extracting on-shell field
irreps of the group will involve interaction. Second, the algebra (5)
can naturally be extended by the new bosonic generator $B^m$ with
commutation relations
\begin{equation}
[B^m,Q_\alpha]=-i{\Gamma^m}_{\alpha\beta}\Sigma^\beta, \qquad
[M_{ab},B_m]=i\eta_{am}B_b-i\eta_{bm}B_a.
\end{equation}
It is straightforward to check that the Jacobi identities hold for the
full algebra. Third, apart from the algebra (5) one can realize the
conjugate superalgebra (see also Ref. 11)
\begin{equation}
\begin{array}{ll} \{Q^\alpha,Q^\beta\}=-2\tilde\Gamma^{m\alpha\beta}P_m,
& [P_m,Q^\alpha]=-i\tilde\Gamma_m{}^{\alpha\beta}\Sigma_\beta,\\
{}[M_{mn},Q^\alpha]=\displaystyle\frac i4 Q^\beta (\Gamma_{mn})_\beta{}
^\alpha, & [M_{mn},\Sigma_\alpha]=-\displaystyle\frac i4
(\Gamma_{mn})_\alpha{}^\beta\Sigma_\beta\end{array}
\end{equation}
with the chirality of the odd generators being opposed to that of the
generators in Eq. (5). It would be interesting to use the doublet (5),
(19) to construct a superparticle model with local $k$-symmetry along
the lines of Ref. 5.

An attractive feature of the Green superalgebra is that it allows one
to formulate the Green--Schwarz superstring in a manifestly
supersymmetric way [11], which was proven to be important in
formulating the theory on the lattice [11]. Let us show that analogous
construction works in terms of the linear transformations (11). Making
use of invariants of the group
\begin{equation}
\begin{array}{c} d\theta^\alpha, \qquad dx^m+i\theta\Gamma^md\theta,\\
d\psi_\alpha-\displaystyle\frac 12 x^m(d\theta\Gamma_m)_\alpha+
dx_m(\theta\Gamma^m)_\alpha+\frac 12(\theta\Gamma^m d\theta)
(\theta\Gamma_m)_\alpha\end{array}
\end{equation}
one can write down a superstring action
\begin{eqnarray}
\lefteqn{S=\displaystyle\int d\tau\,d\sigma\Big\{\frac 12\sqrt{-g}
g^{ij}(\partial_i x^m+i\theta\Gamma^m\partial_i
\theta)(\partial_j x_m+i\theta\Gamma_m\partial_j\theta)+}\cr
&& +\lambda i\epsilon^{ij}(\partial_i\psi
-\displaystyle\frac 12 x_n\partial_i\theta\Gamma^n
+\partial_i x_n\theta\Gamma^n)\partial_j\theta \Big\},
\end{eqnarray}
where $\epsilon^{ij}=-\epsilon^{ji}$, $\epsilon^{01}=-1$ and $\lambda$
is a relative coefficient. Passing to the Hamiltonian formalism one finds
that fermionic constraints of the theory are mixture of half first and
half second class constraints only when\footnote{It is straightforward to
check that two possible values of $\lambda$ in Eq. (22) correspond
to physically
equivalent theories (21). The reason is that the sign in front of the
Wess--Zumino term correlates with the type of two-dimensional projector
$p^{\pm ij}=\frac 12 (g^{ij}\pm\epsilon^{ij}/\sqrt{-g})$ that appears
in the $k$-symmetry transformation law [17].}
\begin{equation}
\lambda =\pm 2/3,
\end{equation}
which, after integrating by parts in Eq.(21), leads to the standard
Green--Schwarz action.

\section{d=11 Bergshoeff--Sezgin superalgebra}
In trying to formulate a supermembrane theory with manifest
supersymmetry and inspired by the $\Gamma$-matrix identities, upon
which the original formulation of super $p$-brane relies [7],
Bergshoeff and Sezgin suggested [2--4] new extensions of the Poincar\'e
superalgebra in $d=11$ (in another respect similar extensions appeared
also in Refs. 18, 19, and 9). The simplest of them reads (our
conventions for $d=11$ are presented in the Appendix)
\begin{eqnarray}
&& \{Q_\alpha,Q_\beta\}=-2{\Gamma^m}_{\alpha\beta}P_m+
(\Gamma_{mn}C)_{\alpha\beta}\Sigma^{mn},\cr
&&{} [Q_\alpha,P_m]=-i(\Gamma_{mn}C)_{\alpha\beta}\Sigma^{n\beta},\\
&& [Q_\alpha,\Sigma^{mn}]=i\Gamma^{[m}{}_{\alpha\beta}
\Sigma^{n]\beta}.\nonumber
\end{eqnarray}
The Jacobi identities for the algebra (23) hold due to the
$\Gamma$-matrix identity
\begin{equation}
{\Gamma^m}_{(\alpha\beta}(\Gamma_{mn}C)_{\gamma\delta)}=0,
\end{equation}
the latter satisfied in $d=4,5,7$ and 11. In this case, with the
standard parametrization of the coset adopted (Eq. (2)), the group-theoretic
construction gives [3]
$$
\begin{array}{l}
\delta_\epsilon\theta^\alpha=\epsilon^\alpha, \qquad
\delta_\epsilon x^m=i\theta\Gamma^m\epsilon,\\
\delta_\epsilon\Phi_{mn}=i(\theta\Gamma_{mn}C\epsilon),\\
\delta_\epsilon\Phi_{n\alpha}=-\displaystyle\frac 12 (\epsilon
\Gamma_{mn}C)_\alpha x^m-\frac 12 (\epsilon\Gamma^m)_\alpha\Phi_{mn}+\\
\qquad+\displaystyle\frac 16 i(\theta\Gamma^m\epsilon)(\theta
\Gamma_{mn}C)_\alpha+\frac 16 i(\theta\Gamma_{mn}C\epsilon)
(\theta\Gamma^m)_\alpha; \\[2ex]
\delta_a x^m=a^m, \qquad \delta_a\Phi_{n\alpha}=\displaystyle\frac 12
a^m(\theta\Gamma_{mn}C)_\alpha;\\
\delta_{\epsilon_{mn}}\Phi_{mn}=\epsilon_{mn}, \qquad
\delta_{\epsilon_{mn}}\Phi_{n\alpha}=\displaystyle\frac 12(\theta
\Gamma^m)_\alpha\epsilon_{mn};\\
\delta_{\epsilon_{n\alpha}}\Phi_{n\alpha}=\epsilon_{n\alpha},\end{array}
\qquad \begin{array}{l} {}\\ (25a)\\ {}\\ {}\\ {}\\ {}\\
(25b) \\ {}\\(25c) \\(25d)
\end{array}
$$
\addtocounter{equation}{1}
where $(x^m,\theta^\alpha,\Phi_{mn},\Phi_{n\alpha})$ are the
coordinates parametrizing the coset and $(a^m,\epsilon^\alpha,
\epsilon_{mn},\epsilon_{n\alpha})$ are the parameters associated to the
generators $(P_m,Q_\alpha,\Sigma^{mn},\Sigma^{n\alpha})$ respectively.
As in the previous case, the $\delta_\epsilon$-transformations
involve nonlinear contributions. However, rewriting the algebra (23) in
the following form:
\begin{equation}
\begin{array}{ll}
{}[\delta_{\epsilon_1},\delta_{\epsilon_2}]=\delta_a+
\delta_{\epsilon_{mn}}, & a^n=2i\epsilon_1\Gamma^n\epsilon_2,\quad
\epsilon_{mn}=2i\epsilon_1\Gamma_{mn}C\epsilon_2;\\
{}[\delta_a,\delta_\epsilon]=\delta_{\epsilon_{n\alpha}}, &
\epsilon_{n\alpha}=-a^m(\epsilon\Gamma_{mn}C)_\alpha;\\
{}[\delta_{\epsilon_{mn}},\delta_\epsilon]=\delta_{\epsilon_{n\alpha}},
& \epsilon_{n\alpha}=-(\epsilon\Gamma^m)_\alpha\epsilon_{mn},\end{array}
\end{equation}
one finds that the linear and nonlinear terms in the variation
$\delta_\epsilon\Phi_{n\alpha}$ make the same contribution into the
first line of Eq. (26)
\begin{equation}
\begin{array}{l} \delta_{\epsilon_1}\Big(-\displaystyle\frac 12 x^m
(\epsilon_2\Gamma_{mn}C)_\alpha-\frac 12 (\epsilon_2\Gamma^m)_\alpha
\Phi_{mn}\Big)-(1\leftrightarrow 2)=\\
\quad =\displaystyle\frac 12 i(\epsilon_1\Gamma_{mn}C\epsilon_2)
(\Gamma^m\theta)_\alpha+\frac 12 i(\epsilon_1\Gamma^m\epsilon_2)
(\theta\Gamma_{mn}C)_\alpha,\\[2ex]
\delta_{\epsilon_1}\Big(\displaystyle\frac 16 i(\theta\Gamma^m
\epsilon_2)(\theta\Gamma_{mn}C)_\alpha+\frac 16 i(\theta\Gamma_{mn}C
\epsilon_2)(\theta\Gamma^m)_\alpha\Big)-(1\leftrightarrow 2)=\\
\quad =\displaystyle\frac 12 i(\epsilon_1\Gamma_{mn}C\epsilon_2)
(\Gamma^m\theta)_\alpha+\frac 12 i(\epsilon_1\Gamma^m\epsilon_2)
(\theta\Gamma_{mn}C)_\alpha.\end{array}
\end{equation}
In checking Eqs. (26) and (27) the consequences of the identity (24)
\begin{equation}
\begin{array}{l}
(\epsilon_2\Gamma_{nm}C)_{(\alpha}(\epsilon_1\Gamma^m)_{\beta)}-
(\epsilon_1\Gamma_{nm}C)_{(\alpha}(\epsilon_2\Gamma^m)_{\beta)}=\\[1ex]
\quad =(\epsilon_1\Gamma^m\epsilon_2)(\Gamma_{nm}C)_{\alpha\beta}+
(\epsilon_1\Gamma_{nm}C\epsilon_2){\Gamma^m}_{\alpha\beta},\\[2ex]
(\epsilon_1\Gamma^m)_\alpha(\theta\Gamma_{mn}C\epsilon_2)+
(\epsilon_1\Gamma_{mn}C)_\alpha(\theta\Gamma^m\epsilon_2)-
(1\leftrightarrow 2)=\\[1ex]
\quad =(\theta\Gamma^m)_\alpha(\epsilon_1\Gamma_{mn}C\epsilon_2)+
(\theta\Gamma_{mn}C)_\alpha(\epsilon_1\Gamma^m\epsilon_2)\end{array}
\end{equation}
are to be used. This observation suggests that one can find another
parametrization of the coset on which Bergshoeff--Sezgin
superalgebra (23) would be linearly realized. The suitable
parametrization looks like
\begin{equation}
g=e^{-ix^nP_n+i\Phi_{n\alpha}\Sigma^{n\alpha}+\frac i2 \Phi_{mn}
\Sigma^{mn}+\frac 23 i\theta^\alpha Q_\alpha} e^{\frac 13 i\theta^\alpha
Q_\alpha}\times SO(1,d-1)
\end{equation}
and the linear version for Eqs. (25a)-(25d) reads
$$
\begin{array}{ll}
\delta_\epsilon\theta^\alpha=\epsilon^\alpha, \qquad \delta_\epsilon
x^m=i\theta\Gamma^m\epsilon, \qquad \delta_\epsilon\Phi_{mn}=
i(\theta\Gamma_{mn}C\epsilon),\\
\delta_\epsilon\Phi_{n\alpha}=-\displaystyle\frac 23 x^m (\epsilon
\Gamma_{mn}C)_\alpha-\frac 23 \Phi_{mn}(\epsilon\Gamma^m)_\alpha; & (30a)\\

\vspace{0.5cm}

\delta_a x^m=a^m, \qquad \delta_a\Phi_{n\alpha}=\displaystyle\frac 13
a^m(\theta\Gamma_{mn}C)_\alpha; & (30b)\\

\vspace{0.3cm}

\delta_{\epsilon_{mn}}\Phi_{mn}=\epsilon_{mn}, \qquad
\delta_{\epsilon_{mn}}\Phi_{n\alpha}=\displaystyle\frac 13 \epsilon_{mn}
(\theta\Gamma^m)_\alpha; & (30c)\\

\vspace{0.3cm}

\delta_{\epsilon_{n\alpha}}\Phi_{n\alpha}=\epsilon_{n\alpha}; &
(30d)\end{array}
$$
\addtocounter{equation}{1}
As already might be expected, the generators of the nonlinear
transformations (25)
\begin{equation}
\begin{array}{l}
Q_\alpha=i\partial_\alpha+(\theta\Gamma^n)_\alpha\partial_n +
(\theta\Gamma_{mn}C)_\alpha\partial^{mn}-\Big(\displaystyle\frac i2
x^m(\Gamma_{mn}C)_{\alpha\beta}+\\
\qquad+\displaystyle\frac i2 {\Gamma^m}_{\alpha\beta}\Phi_{mn}-\frac 16
(\theta\Gamma^m)_\alpha(\theta\Gamma_{mn}C)_\beta-\frac 16 (\theta
\Gamma_{mn}C)_\alpha(\theta\Gamma^m)_\beta\Big)\partial^{n\beta},\\
P_m=-i\partial_m-\displaystyle\frac i2 (\theta\Gamma_{mn}C)_\alpha
\partial^{n\alpha},\\
\Sigma^{mn}=i\partial^{[mn]}+\displaystyle\frac i2(\theta\Gamma^{[m}
)_\alpha\partial^{n]\alpha}\\
\Sigma^{n\alpha}=i\partial^{n\alpha}\end{array}
\end{equation}
and ones of the linear transformations (30)
\begin{equation}
\begin{array}{l}
Q_\alpha=i\partial_\alpha+(\theta\Gamma^n)_\alpha\partial_n +
(\theta\Gamma_{mn}C)_\alpha\partial^{mn}-\\
\quad-\Big(\displaystyle\frac 23 ix^m(\Gamma_{mn}C)_{\alpha\beta}+
\frac 23 i\Phi_{mn}{\Gamma^m}_{\alpha\beta}\Big)\partial^{n\beta},\\
P_m=-i\partial_m-\displaystyle\frac i3 (\theta\Gamma_{mn}C)_\alpha
\partial^{n\alpha},\\
\Sigma^{mn}=i\partial^{[mn]}+\displaystyle\frac i3(\theta\Gamma^{[m}
)_\alpha\partial^{n]\alpha},\\
\Sigma^{n\alpha}=i\partial^{n\alpha}\end{array}
\end{equation}
belong to equivalent representations of the superalgebra (23)
\begin{equation}
T_{i\rm(nonlinear)}=ST_{i\rm(linear)}S^{-1},
\end{equation}
where\footnote{A comparison of the parametrizations (2) and (29) with
the latter rewritten in the form
\begin{eqnarray*}
& g=e^{-ix^nP_n+i\Phi_{n\alpha}\Sigma^{n\alpha}+\frac i2\Phi_{mn}
\Sigma^{mn}+\frac 23 i\theta^\alpha Q_\alpha}e^{\frac 13 i\theta^\alpha
Q_\alpha}\times SO(1,d-1)= \\
& =e^{-ix^nP_n+i\theta^\alpha Q_\alpha+\frac i2\Phi_{mn}\Sigma^{mn}
+i\left(\Phi_{n\alpha} +\frac 16 x^m(\theta\Gamma_{mn}C)_\alpha
+\frac 16\Phi_{mn}(\theta\Gamma^m)_\alpha\right)\Sigma^{n\alpha}},
\end{eqnarray*}
suggests the explicit form for the operator $S$.}
\begin{equation}
S=e^{-\frac 16 \left(x^m(\theta\Gamma_{mn}C)_\alpha+\Phi_{mn}
(\theta\Gamma^m)_\alpha\right)\partial^{n\alpha}}.
\end{equation}
In Eqs. (31)--(34) we denoted $\partial/\partial\theta^\alpha=
\partial_\alpha$, $\partial/\partial x^n=\partial_n$,
$\partial/\partial\Phi_{n\alpha}=\partial^{n\alpha}$,
$\partial/\partial\Phi_{mn}=\partial^{mn}$ and set $T_i\equiv
(Q_\alpha,P^m,\Sigma^{mn},\Sigma^{n\alpha})$.

\section{Inclusion of $\Sigma^{\alpha\beta}$-charge}
Close examination
of invariants of the group (25) (or (30)) shows that they
are not sufficient to construct a supermembrane theory with local
$k$-symmetry. Further extension by $\Sigma^{\alpha\beta}$-generator was
proven [3] to be necessary. The structure relations of the algebra have
the form\footnote{Recently, an extension of Eq. (35) by super five-form
charge has been proposed [4]. The $M$-algebra may suggest an attractive
way to construct a super five-brane theory in eleven dimensions. The
possibility to realize the $M$-algebra by linear transformations will
be examined elsewhere.} [3]
\begin{equation}
\begin{array}{l}
\{Q_\alpha,Q_\beta\}=-2{\Gamma^m}_{\alpha\beta}P_m+(\Gamma_{mn}C
)_{\alpha\beta}\Sigma^{mn},\\
{}[Q_\alpha,P_m]=-i(\Gamma_{mn}C)_{\alpha\beta}\Sigma^{n\beta},\\
{}[P_m,P_n]=i(\Gamma_{mn}C)_{\alpha\beta}\Sigma^{\alpha\beta},\\
{}[P_m,\Sigma^{np}]=-\displaystyle\frac 12 i{\delta_m}^{[n}
\Gamma^{p]}{}_{\alpha\beta}\Sigma^{\alpha\beta},\\[2ex]
{}[Q_\alpha,\Sigma^{mn}]=i \Gamma^{[m}{}_{\alpha\beta}
\Sigma^{n]\beta},\\
\{Q_\alpha,\Sigma^{n\beta}\}=\Big(\displaystyle\frac 12
{\Gamma^n}_{\gamma\delta}{\delta_\alpha}^\beta+
4{\Gamma^n}_{\gamma\alpha}{\delta_\delta}^\beta\Big)
\Sigma^{\gamma\delta}.\end{array}
\end{equation}
The Jacobi identities for (35) hold due to Eq. (24). As compared to the
usual Poincar\'e superalgebra, one finds the super two-form charge
$\Sigma^{AB}=(\Sigma^{mn},\Sigma^{m\alpha},\Sigma^{\alpha\beta})$ in
Eq. (35) which was connected [3] with the existence of supermembrane
solution [20] of $d=11$ supergravity.

Although in this case the commutation relations look rather
complicated, the modified algebra (35) can be linearly realized like
its contraction (23). Omitting details we present here the final result.

The transformations on the coset space (with the standard
parametrization (2) adopted) are given by Eq. (25) with the following
transformation laws for the coordinates $\Phi_{\alpha\beta}$ associated
to the new generators $\Sigma^{\alpha\beta}$ added (compare with Ref. 3
where another parametrization of the coset has been chosen)
$$
\begin{array}{ll}
\delta_\epsilon\Phi_{\alpha\beta}=\displaystyle\frac i2 (\Phi_{m\gamma}
\epsilon^\gamma){\Gamma^m}_{\alpha\beta}+2i\Phi_{m(\alpha}
(\epsilon\Gamma^m)_{\beta)}-\\
\quad-\displaystyle\frac 14\Phi_{nm}i(\theta\Gamma^n\epsilon)
{\Gamma^m}_{\alpha\beta}+\frac 13 x^ni(\epsilon\Gamma_{nm}C)_{(\alpha}
(\theta\Gamma^m)_{\beta)}+\\
\quad+\displaystyle\frac 13 \Phi_{nm}i(\epsilon\Gamma^n)_{(\alpha}
(\theta\Gamma^m)_{\beta)}-\frac 13 x^ni(\theta\Gamma^m\epsilon)
(\Gamma_{mn}C)_{\alpha\beta}+\\
\quad+\displaystyle\frac 1{12}x^n i(\theta\Gamma_{nm}C\epsilon)
{\Gamma^m}_{\alpha\beta}; & (36a)\\
\delta_a\Phi_{\alpha\beta}=-a^nx^m(\Gamma_{nm}C)_{\alpha\beta}-
\displaystyle\frac 12 a^m\Phi_{mn}{\Gamma^n}_{\alpha\beta}-\\
\quad-\displaystyle\frac 13 a^ni(\theta\Gamma_{nm}C)_{(\alpha}
(\theta\Gamma^m)_{\beta)}; & (36b)\end{array}
$$

$$
\begin{array}{ll}
\delta_{\epsilon_{mn}}\Phi_{\alpha\beta}=\displaystyle\frac 12 x^n
\epsilon_{nm}{\Gamma^m}_{\alpha\beta}-\frac 13\epsilon_{nm}i
(\theta\Gamma^n)_{(\alpha}(\theta\Gamma^m)_{\beta)}; & (36c)\\
\delta_{\epsilon_{n\alpha}}\Phi_{\alpha\beta}=\displaystyle\frac 12 i
(\theta^\gamma\epsilon_{n\gamma}){\Gamma^n}_{\alpha\beta}
-2i\epsilon_{n(\alpha}(\theta\Gamma^n)_{\beta)}; & (36d) \\
\delta_{\epsilon_{\alpha\beta}}\Phi_{\alpha\beta}=
\epsilon_{\alpha\beta}. & (36e)\end{array}
$$
Taking the linearization for Eq. (25) from Eq. (30) and supplementing
it with
$$
\begin{array}{ll}
\delta_\epsilon\Phi_{\alpha\beta}=3i(\Phi_{n\gamma}\epsilon^\gamma)
{\Gamma^n}_{\alpha\beta}+3i\Phi_{n(\alpha}(\epsilon\Gamma^n)_{\beta)};
& (37a)\\
\delta_a\Phi_{\alpha\beta}=-2a^m\Phi_{mn}{\Gamma^n}_{\alpha\beta}+
a^mx^n(\Gamma_{nm}C)_{\alpha\beta}; & (37b)\\
\delta_{\epsilon_{mn}}\Phi_{\alpha\beta}=-\epsilon_{mn}x^m
{\Gamma^n}_{\alpha\beta}; & (37c)\\
\delta_{\epsilon_{n\alpha}}\Phi_{\alpha\beta}=2i(\epsilon_{n\gamma}
\theta^\gamma){\Gamma^n}_{\alpha\beta}-i\epsilon_{n(\alpha}
(\theta\Gamma^n)_{\beta)}; & (37d)\\
\delta_{\epsilon_{\alpha\beta}}\Phi_{\alpha\beta}=\epsilon_{\alpha\beta},
& (37e)\end{array}
$$
one gets the linear version for Bergshoeff--Sezgin superalgebra
(35). Making use of Eq. (28), it is straightforward to check that
transformations (30), (37) satisfy the algebra
\addtocounter{equation}{2}
\begin{equation}
\begin{array}{ll}
{}[\delta_{\epsilon_1},\delta_{\epsilon_2}]=\delta_a+
\delta_{\epsilon_{mn}}, & a^m=2i\epsilon_1\Gamma^m\epsilon_2, \qquad
\epsilon_{mn}=2i(\epsilon_1\Gamma_{mn}C\epsilon_2);\\
{}[\delta_a,\delta_\epsilon]=\delta_{\epsilon_{n\alpha}}, &
\epsilon_{n\alpha}=-a^m(\epsilon\Gamma_{mn}C)_\alpha;\\
{}[\delta_{\epsilon_{mn}},\delta_\epsilon]=\delta_{\epsilon_{n\alpha}},
& \epsilon_{n\alpha}=-\epsilon_{mn}(\Gamma^m\epsilon)_\alpha;\\
{}[\delta_\epsilon,\delta_{\epsilon_{n\alpha}}]=
\delta_{\epsilon_{\alpha\beta}}, & \epsilon_{\alpha\beta}=
-i(\epsilon_{n\gamma}\epsilon^\gamma){\Gamma^n}_{\alpha\beta}
-4i\epsilon_{n(\alpha}(\Gamma^n\epsilon)_{\beta)};\\
{}[\delta_{a_1},\delta_{a_2}]=\delta_{\epsilon_{\alpha\beta}}, &
\epsilon_{\alpha\beta}=2{a_2}^m{a_1}^n(\Gamma_{nm}C)_{\alpha\beta};\\
{}[\delta_a,\delta_{\epsilon_{mn}}]=\delta_{\epsilon_{\alpha
\beta}}, & \epsilon_{\alpha\beta}=a^m\epsilon_{mn}
{\Gamma^n}_{\alpha\beta},\end{array}
\end{equation}
which, being rewritten in terms of the generators, coincides with Eq. (35).

\bigskip

Thus, in this work we have demonstrated that nontrivial extensions of
the Poincar\'e superalgebra in higher dimensions can be realized
without spoiling the linear structure of the original super Poincar\'e
transformations. We hope that analogous construction will work in the
case of the $M$-algebra [4] which may suggest considerable
simplification in constructing a super five-brane theory in eleven
dimensions.

\bigskip

\vspace{0.2cm}

{\bf Acknowledgments}

\vspace{0.2cm}

One of the authors (A.G.) thanks V.P. Akulov, E.A. Ivanov, D.P. Sorokin and
M.A. Vasiliev for valuable discussions. The hospitality extended to him
at the ICTP is also gratefully acknowledged. Research was supported by
Joint DFG-RFBR project No 96-02-00180G (A.D.) and by INTAS Grant
No 93-2492-ext (within the research program of ICFPM),
INTAS-RFBR Grant No 95-829 (A.G.).

\setcounter{equation}{0}
\def\theequation{A.\arabic{equation}}

\vspace{0.2cm}

{\bf Appendix}

\vspace{0.2cm}

In this Appendix we present the spinor notations adopted
for $d=11$ in this work.

A conventional way to build $\Gamma$-matrices in eleven dimensions is
to use those of ten-dimensional space-time. The imaginary (Majorana)
representation
\begin{equation}
{\Gamma^M}_\alpha{}^\beta=i\left(\begin{array}{cc} 0 & {\Gamma^M}_{AB}\\
\tilde\Gamma^{M\,AB} & 0\end{array}\right), \qquad \alpha=1,\dots,32;
\quad M=0,\dots,9,
\end{equation}
is of common use in $d=10$. Here ${\Gamma^M}_{AB}$,
$\tilde\Gamma^{M\,AB}$ are $16\times 16$ matrices which form ``chiral''
representation. They are real and symmetric, obeying the algebra
\begin{equation}
\Gamma^M\tilde\Gamma^N+\Gamma^N\tilde\Gamma^M=-2\eta^{MN}, \qquad
\eta^{MN}=(+,-,-,\dots).
\end{equation}
The explicit form for the matrices is
\begin{eqnarray}
& \Gamma^0=\left(\begin{array}{cc} {\bf 1}_8 & 0\\ 0 & {\bf 1}_8
\end{array}\right), & \tilde\Gamma^0=\left(\begin{array}{cc}
-{\bf 1}_8 & 0\\ 0 & -{\bf 1}_8\end{array}\right), \cr
& \Gamma^i=\left(\begin{array}{cc} 0 & {\gamma^i}_{a\dot a}
\\ {\gamma^i}_{\dot aa} & 0 \end{array}\right), &
\tilde\Gamma^i=\left(\begin{array}{cc} 0 & {\gamma^i}_{a\dot a}
\\ {\gamma^i}_{\dot aa} & 0 \end{array}\right), \\
& \Gamma^9=\left(\begin{array}{cc} {\bf 1}_8 & 0\\ 0 & -{\bf 1}_8
\end{array}\right), & \tilde\Gamma^9=\left(\begin{array}{cc}
{\bf 1}_8 & 0\\ 0 & -{\bf 1}_8\end{array}\right), \nonumber
\end{eqnarray}
where ${\gamma^i}_{a\dot a}$, ${\gamma^i}_{\dot aa}\equiv
({\gamma^i}_{a\dot a})^{\rm T}$ are $SO(8)$ $\gamma$-matrices [17]
\begin{equation}
{\gamma^i}_{a\dot a}{\gamma^j}_{\dot ab}+{\gamma^j}_{a\dot a}
{\gamma^i}_{\dot ab}=2\delta^{ij}\delta_{ab}, \quad i=1,\dots,8; \quad
a,\dot a=1,\dots,8.
\end{equation}
The properties of $\Gamma$, $\tilde\Gamma$ induce the relations for
${\Gamma^M}_\alpha{}^\beta$
\begin{equation}
\begin{array}{c} \Gamma^M\Gamma^N+\Gamma^N\Gamma^M=2\eta^{MN}, \qquad
\eta^{MN}=(+,-,-,\dots),\\
(\Gamma^M)^*=-\Gamma^M, \qquad (\Gamma^0)^{\rm T}=-\Gamma^0, \qquad
(\Gamma^i)^{\rm T}=\Gamma^i, \quad i=1,\dots,9.\end{array}
\end{equation}
The product $\Gamma^0\Gamma^1\cdots\Gamma^9$ is known as $\Gamma^{11}$
\begin{eqnarray}
& \Gamma^{11}=\left(\begin{array}{cc} {\bf 1}_{16} & 0\\
0 & -{\bf 1}_{16}\end{array}\right), \\
& (\Gamma^{11})^2={\bf 1}_{32}, \qquad
\Gamma^{11}\Gamma^M=-\Gamma^M\Gamma^{11}, \nonumber
\end{eqnarray}
and serves to extract two inequivalent irreducible spinor representations
of the Lorentz group in $d=10$ (right-handed and left-handed Weyl
spinors: $\Psi_{R,L}=\frac 12(1\pm\Gamma^{11})\Psi$ with $\Psi$ a Dirac
spinor).

With the $\Gamma^{11}$ at hand, construction of $d=11$
$\Gamma$-matrices present no special problem [21]. It is easy to check
that the set
\begin{equation}
{\Gamma^m}_\alpha{}^\beta\equiv[{\Gamma^M}_\alpha{}^\beta,
i{\Gamma^{11}}_\alpha{}^\beta], \qquad m=0,\dots,10; \quad
\alpha=1,\dots,32,
\end{equation}
satisfies the needed algebra
\begin{equation}
\Gamma^m\Gamma^n+\Gamma^n\Gamma^m=2\eta^{mn}, \qquad
\eta^{mn}=(+,-,-,\dots)
\end{equation}
and possesses the properties
\begin{equation}
(\Gamma^m)^*=-\Gamma^m, \qquad (\Gamma^0)^{\rm T}=-\Gamma^0, \qquad
(\Gamma^i)^{\rm T}=\Gamma^i, \quad i=1,\dots,10.
\end{equation}

Under action of the Lorentz group a $d=11$ Dirac spinor is transformed
as
\begin{equation}
\delta\Psi_\alpha=\displaystyle\frac 18 \omega^{mn}
(\Gamma_{mn}\Psi)_\alpha,\qquad
\Gamma_{mn}\equiv\Gamma_m\Gamma_n-\Gamma_n\Gamma_m.
\end{equation}
In the imaginary representation (A.7)--(A.9) the reality condition
\begin{equation}
\Psi^*_\alpha=\Psi_\alpha
\end{equation}
is compatible with Eq. (A.10) which defines a Majorana spinor.

In studying the extended superalgebras in eleven dimensions it is
convenient to deal with symmetric $\Gamma$-matrices much as the chiral
representation (A.2), (A.3) is of common use in $d=10$. It is
straightforward to check that the matrices
\begin{equation}
\begin{array}{l}
{\Gamma^m}_{\alpha\beta}\equiv\left[\left(\begin{array}{cc}
-\tilde\Gamma^{M\,AB} & 0\\ 0 & {\Gamma^M}_{AB}\end{array}\right),
\left(\begin{array}{cc} 0 & {\bf 1}_{16}\\ {\bf 1}_{16} & 0\end{array}
\right)\right],\\
\tilde\Gamma^{m\,\alpha\beta}\equiv\left[\left(\begin{array}{cc}
{\Gamma^M}_{AB} & 0\\ 0 & -\tilde\Gamma^{M\,AB}\end{array}\right),
\left(\begin{array}{cc} 0 & -{\bf 1}_{16}\\ -{\bf 1}_{16} & 0\end{array}
\right)\right]\end{array}
\end{equation}
satisfy the algebra
\begin{equation}
\Gamma^m\tilde\Gamma^n+\Gamma^n\tilde\Gamma^m=2\eta^{mn}.
\end{equation}
Obviously, they are real and symmetric. In these terms one can define
two spinor representations
\begin{equation}
\begin{array}{ll}
\delta\Psi_\alpha=\displaystyle\frac 18\omega^{mn} (\Gamma_{mn})_\alpha
{}^\beta\Psi_\beta, &
\Gamma_{mn}\equiv\Gamma_m\tilde\Gamma_n-\Gamma_n\tilde\Gamma_m,\\
\delta\chi^\alpha=\displaystyle\frac 18\omega^{mn}
(\tilde\Gamma_{mn})^\alpha{}_\beta\chi^\beta, &
\tilde\Gamma_{mn}\equiv\tilde\Gamma_m\Gamma_n-\tilde\Gamma_n\Gamma_m,
\end{array} \end{equation}
which, however, are equivalent. The Lorentz invariant charge
conjugation matrix
\begin{eqnarray}
& C_{\alpha\beta}=i\left(\begin{array}{cc} 0 & {\bf 1}_{16}\\
-{\bf 1}_{16} & 0\end{array}\right), \\
& (\Gamma_{mn})_\alpha{}^\beta C_{\beta\gamma}=C_{\alpha\beta}
(\tilde\Gamma_{mn})^\beta{}_\gamma \nonumber
\end{eqnarray}
can be used to raise or lower spinor indices. One can check also,
that the following relations:
\begin{equation}
\begin{array}{c} (\Gamma_{mn}C)_{\alpha\beta}=(\Gamma_{mn}
C)_{\beta\alpha},\\
\tilde\Gamma^m=C\Gamma^mC^{-1}\end{array}
\end{equation}
hold.


\begin{thebibliography}{nn}
\bibitem{1} M.B. Green, Phys. Lett. B {\bf 223} (1989) 157.
\bibitem{2} E. Bergshoeff and E. Sezgin, Phys. Lett. B {\bf 232} (1989)
96.
\bibitem{3} E. Bergshoeff and E. Sezgin, Phys. Lett. B {\bf 354} (1995)
256; hep-th/9504140; 9512082.
\bibitem{4} E. Sezgin, hep-th/9609086.
\bibitem{5} A.A. Deriglazov and A.V. Galajinsky, Mod. Phys. Lett. A
{\bf 9} (1994) 3445.
\bibitem{6} J. Hughes, J. Liu, and J. Polchinski, Phys. Lett. B {\bf
180} (1986) 370.
\bibitem{7} E. Bergshoeff, E. Sezgin, and P.K. Townsend, Phys. Lett. B
{\bf 189} (1987) 75.
\bibitem{8} A. Ach\'ucarro, J.M. Evans, P.K. Townsend, and D.L.
Wiltshire, Phys. Lett. B {\bf 198} (1987) 441.
\bibitem{9} J.A. de Azc\'arraga, J.P. Gauntlett, J.M. Izquierdo, and P.K.
Townsend, Phys. Rev. Lett. {\bf 63} (1989) 2443.
\bibitem{10} J.A. de Azc\'arraga and J. Lukierski, Phys. Lett. B {\bf
113} (1982) 170; Phys. Rev. D {\bf 38} (1988) 509;\\
J.M. Evans, Nucl. Phys. B {\bf 331} (1990) 711; \\
A.A. Deriglazov, A.V. Galajinsky, and S.L. Lyakhovich, Nucl. Phys. B
{\bf 473} (1996) 245.
\bibitem{11} W. Siegel, Phys. Rev. D {\bf 50} (1994) 2799.
\bibitem{12} S. Helgason, {\it Differential Geometry and Symmetric Spaces},
(Academic, New York 1962);\\
V.I. Ogievetsky, Non-linear realizations of internal and
spacetime symmetries, in: Proc. 10th Karpacz Winter School of
Theoretical Physics (1974).
\bibitem{13}
E.S. Fradkin and M.A. Vasiliev, Phys. Lett. B {\bf 189} (1987) 89;
Nucl. Phys. B {\bf 291} (1987) 141;\\
M.A. Vasiliev, Phys. Lett. B {\bf 285} (1992) 225.
\bibitem{14} W. Siegel, Nucl. Phys. B {\bf 263} (1985) 285.
\bibitem{15} A.A. Deriglazov and A.V. Galajinsky, Phys. Rev. D {\bf 54}
(1996) 5195.
\bibitem{16} E.A. Ivanov and A.S. Sorin, J. Phys. A {\bf 13} (1980) 1159.
\bibitem{17} M.B. Green, J.H. Schwarz, and E. Witten, {\it Superstring
Theory} (Cambridge University Press, Cambridge 1987).
\bibitem{18} R. D'Auria and P. Fr\'e, Nucl. Phys. B {\bf 201} (1982)
101.
\bibitem{19} J.W. van Holten and A. van Proeyen, J. Phys. A {\bf 15}
(1982) 763.
\bibitem{20} M.J. Duff and K.S. Stelle, Phys. Lett. B {\bf 253} (1991)
113.
\bibitem{21} J. Scherk, Extended supersymmetry in extended supergravity
theory, in: {\it Recent Developments in Gravitation}, eds. M. Levy and
S. Deser, Plenum Publ. Corp. (1979).
\end{thebibliography}
\end{document}